\documentclass[11pt,a4paper]{article}
\usepackage{epsfig,graphicx,amsfonts,amsmath}

\usepackage{cite}
\RequirePackage{mathrsfs}
\newlength{\dinwidth}
\newlength{\dinmargin}
\def\eq#1{{Eq.~(\ref{#1})}}
\newcommand{\Le}{\left(}
\newcommand{\Ra}{\right)}

\newcommand{\beq}{\begin{equation}}
\newcommand{\eeq}{\end{equation}}
\newcommand{\beqar}{\begin{eqnarray}}
\newcommand{\eeqar}{\end{eqnarray}}
\newcommand{\D}{\partial}

\newcommand{\tv}{\textsl{v}}

\newcommand{\T}{{\cal T}}
\date{}
\begin{document}
%\input{header.tex}
%\begin{flushright}
%\vspace{-0.5cm}
%{\Large \bf DRAFt}\\
%\today
%\end{flushright}
%\thispagestyle{empty}

\title {{~}\\
{\Large \bf High energy QCD Lipatov's effective action in Euclidean space }}
%\author{
%{~}\\
%{\large \,\,
%}\\[7mm]
%{\it\normalsize  $^{(1) }$Physics Department, Ariel University, Ariel 40700, Israel}\\}
\author{
{~}\\
{\large
%J.~Bartels$^{(1)}$,
S.~Bondarenko$^{(1) }$
%S.~Pozdnyakov$^{(1) }$
}\\[7mm]
{\it\normalsize  $^{(1) }$ Physics Department, Ariel University, Ariel 40700, Israel}\\
}

\maketitle
\thispagestyle{empty}

\begin{abstract}

 The continuation of high energy QCD Lipatov's effective action to Euclidean space is performed. The resulting Euclidean QCD RFT action
is considered separately in Euclidean "light-cone" coordinates and axial gauge suitable for the numerical and analytical calculations correspondingly.
The further application of the obtained results is also discussed.

\end{abstract}

\section{Introduction}

  Lipatov's effective action approach, \cite{LipatovEff,LipatovEff1, Our1,Our2,Our3,Our4,Our41,Our5,Our6,Our7}, can be considered as Regge Field Theory (RFT) constructed  on the base of 
QCD and intended to to take account of unitarity corrections to the high-energy scattering amplitudes in the multi-Regge kinematics, see \cite{EffAct,Nefedov}. It also can be considered as a further generalization of phenomenological RFT \cite{Gribov,0dim1,0dim2,0dim3,0dim4}. The formalism is based on the reggeized gluons (reggeons) as the main degrees of freedom, 
with the Pomeron calculus, \cite{Pom1,Pom2,Pom22,Pom3,Pom4}, introduced on the base of colorless Reggeons states.
	
 The transition to QCD RFT formulation of the theory is performed
with the introduction of the generating functional for the reggeized gluons fields $A_{\pm}$ obtained by an integration out of the gluon fields $v$ from the
the $S_{eff}[v,\,A]$:
\beq\label{Add1}
e^{\imath\,\Gamma[A]}\,=\,\int\,D v\,e^{\imath\,S_{eff}[v,\,A\,]}
\eeq
with Lipatov's effective action defined as QCD action plus effective currents for the longitudinal gluon fields:
\beq\label{Add2}
S_{eff}\,=\,-\,\int\,d^{4}\,x\,\Le\,\frac{1}{4}\,G_{\mu \nu}^{a}\,G^{\mu \nu}_{a}\, \,+\,tr\,\left[\,\Le\,\T_{+}(v_{+})\,-\,A_{+}\,\Ra\,j_{reg}^{+}\,+\,
\Le\,\T_{-}(\textsl{v}_{-})\,-\,A_{-}\,\Ra\,j_{reg}^{-}\,\right]\,\Ra\,
\eeq
and effective currents defined as
\beq\label{Add3}
\T_{\pm}(v_{\pm})\,=\,\frac{1}{g}\,\D_{\pm}\,O(v_{\pm})\,=\,
v_{\pm}\,O(v_{\pm})\,,\,\,\,\,
j_{reg\,a}^{\pm}\,=\,\frac{1}{C(R)}\,\D_{i}^{2}\,A_{a}^{\pm}\,,
\eeq
where  $C(R)$ is eigenvalue of Casimir operator in the representation R, $tr(T^{a} T^{b})\,=\,C(R)\,\delta^{a b}$\, 
$A_{\pm}$ as Reggeon fields, see \cite{LipatovEff,LipatovEff1,Our1,Our2}.
The form of the Lipatov's operator  $O$ (and correspondingly $\T$) depends on the particular process of interests, see \cite{Our4},
the simplest choice of the operators are Wilson lines (ordered
exponential) of the longitudinal gluon fields:
\beq\label{Add4}
O(v_{\pm})\,=\,P\,e^{g\,\int_{-\infty}^{x^{\pm}}\,d x^{\pm}\,v_{\pm}(x^{+},\,x^{-},\,x_{\bot})}\,,\,\,\,\,\tv_{\pm}\,=\,\imath\,T^{a}\,v_{\pm}^{a}\,,
\eeq
see also \cite{Nefedov}. 
There are additional kinematical constraints for the Reggeon fields
\beq\label{Ef4}
\partial_{-}\,A_{+}\,=\,\partial_{+}\,A_{-}\,=\,0\,,
\eeq
corresponding to the strong-ordering of the Sudakov components in the multi-Regge kinematics,
see \cite{LipatovEff,LipatovEff1,Our4}. The action is constructed by the request that the  LO value of the classical gluon fields in the solutions of equations of motion will be fixed as
\beq\label{Ef8}
\textsl{v}_{\pm}^{cl}\,=\,A_{\pm}\,,
\eeq
this condition also can be considered as the definition of the Reggeon fields.
The further use of the \eq{Add2} action is based on the different perturbative calculation schemes, see \cite{Our41,Our5,Our6,Our7,EffAct,Nefedov}, of course all the calculations are performed in the Minkowski space. 

 The problem of the continuation of the action to the Euclidean space, respectively, is important task due to the two interesting issues. 
The fist one is the formulation of the action for the possible lattice (numerical) calculation of any objects of interests in the Euclidean space.
The second one is the analytical investigation of the non-perturbative contributions, such as instantons for example, to the Reggeons interactions vertices and amplitudes.
There are many calculations which have be done in this direction, see for example \cite{Inst}, but the Lipatov's action continued to Euclidean space can serve as some generic approach for the consistent account of the instanton-Reggeon interactions and instanton contributions to the high-energy scattering amplitudes.
Therefore, the paper is organized as follows. In the next section we introduce all required  tools and notions for the continuation of the action to Euclidean space.
In the Section 3 we consider the continuation of the Lipatov's action in the Euclidean space. Section 4 and 5 are about the different forms of the action. In Section 4 we discuss the action written in the terms of 
new coordinates similar to the light-cone coordinates of Minkowski space, we will call it further as Euclidean "light-cone" coordinates. 
In the Section 5 we consider the axial gauge for the action which is suitable for the
analytical calculations in the framework. The last section is Conclusion where the further applications of the results are discussed.

\section{Euclidean space kinematics}

  The main task of the problem is the continuation of the effective currents of the action to Euclidean space. 
We begin from the usual four-momentum vectors in Minkowski space used for the description of the high energy scattering kinematics:  
\beq\label{MS1}
p_{\,1}^{\,\mu}\,=\,\frac{\sqrt{s}}{2}(1,1,0_{\bot}),\,\,\,p_{\,2}^{\,\mu}\,=\,\frac{\sqrt{s}}{2}(1,-1,0_{\bot})\,,
\eeq
where 
%\beq\label{MS2}
%s\,=\,E^2\,.
%\eeq
%Kinematics:
\beq\label{MS3}
s\,=\,(p_{1}\,+\,p_{2})^2\,,\,\,\,\,t\,=\,(p_{1}\,-\,p_{1}^{'})^2\,,\,\,\,\,s\,=\,(p_{1}\,-\,p_{2}^{'})^2\,.
\eeq
%For masless particles:
%\beq\label{MS4}
%s\,=\,2\,p_{1}\,p_{2}\,=\,E^{2}\,.
%\eeq
In the light cone coordinate frame these vectors can be rewritten as
\beq\label{MS101}
p_{\,1\,L.C.}^{\,\mu}\,=\,\sqrt{\frac{s}{2}}(1,\beta,0_{\bot}),\,\,\,p_{\,2\,L.C.}^{\,\mu}\,\,=\,\sqrt{\frac{s}{2}}(\beta,1,0_{\bot})
\eeq
with $\beta$ as some regularization introduced in order to regularize the rapidity divergences of the corresponding integrals, see \cite{BalOp} for example.
Application of these vectors to the \eq{Add3} current terms provides
\beq\label{MS102}
\sqrt{\frac{2}{s}}\,p_{1\,L.C}^{\,\mu}\,\D_{\mu}\,\stackrel{\beta \rightarrow 0}{\rightarrow}\,\D_{+}\,,\,\,\,
\sqrt{\frac{2}{s}}\,p_{2\,L.C.}^{\,\mu}\,\D_{\mu}\,\stackrel{\beta \rightarrow 0}{\rightarrow}\,\D_{-}\
\eeq
as requested. Therefore, the Eikonal interaction term introduced in \cite{Our4,Our5}, see  also\eq{ES1} below, can be written as following:
\beq\label{MS103}
\frac{2}{s}\,\Le \, p_{\,1\,L.C.}^{\,\mu}\,\D_{\mu}\,O(\tv_+)\Ra\,\D_{\bot}^{2}\, \Le \, p_{\,2\,L.C.}^{\,\nu}\,\D_{\nu}\,O(\tv_-)\Ra\,
\eeq
which is fully covariant, i.e. it can be also in the usual Minkowski coordinates as
\beq\label{MS104}
\frac{2}{s}\,\Le \, p_{\,1}^{\,\mu}\,\D_{\mu}\,O(\tv_+)\Ra\,\D_{\bot}^{2}\, \Le \, p_{\,2}^{\,\nu}\,\D_{\nu}\,O(\tv_-)\Ra\,,
\eeq
that can considered as a step to the continuation of the effective action to Euclidean space. 
Nevertheless, it is not easy to work with $\beta$ as parameter performing the continuation for the arbitrary angle between particles trajectories in the Euclidean space, 
see \cite{Megg} . Therefore, considering the hyperbolic angle between the light cone directions in the Minkowski space, 
we introduce the following vectors of the directions of the relativistic particles motion in the Minkowski space:
\beq\label{MS105}
n_{\,1}^{\,\mu}\,=\,\frac{1}{\sqrt{2}}\,(1,\tanh(\gamma / 2),0_{\bot}),\,\,\,n_{\,2}^{\,\mu}\,=\,\frac{1}{\sqrt{2}}\,(1,-\tanh(\gamma / 2),0_{\bot}),
\eeq  
with the following form of the same vectors in the light cone coordinates:
\beq\label{MS11}
n_{\,+\,L.C.}^{\,\mu}\,=\,\frac{1}{2}\,(1+\tanh(\gamma / 2),1-\tanh(\gamma / 2),0_{\bot}),\,\,\,
n_{\,-\,L.C.}^{\,\mu}\,=\,\frac{1}{2}\,(1-\tanh(\gamma / 2),1+\tanh(\gamma / 2),0_{\bot})\,.
\eeq
As usual, for the
\beq\label{MS111}
p^{2}\,=\,m^2
\eeq
we have at high energy
\beq\label{MS112}
\gamma\,\approx\,\ln(s/m^2)\,,\,\,\,\beta\,=\,m^2 / s\,.
\eeq
The Wick rotation of the vectors to the Euclidean space can be done now by
\beq\label{MS14}
\gamma\,\rightarrow\,2 \imath \phi
\eeq
continuation, where $2\phi$ is an angle between trajectories of the two particles in the c.m.f. in Euclidean space.
We obtain correspondingly for \eq{MS105}:
%\beqar\label{MS15}
%&\,&
\beq \label{MS15}
%&\,&
n_{\,1\,E}^{\,\mu}\,=\,\frac{1}{\sqrt{2}}\,(1,\imath \tan(\phi) ,0_{\bot}),\,\,\,
n_{\,2\,E}^{\,\mu}\,=\,\frac{1}{\sqrt{2}}\,(1 , -\imath\tan(\phi) ,0_{\bot})\,, 
\eeq
%&\,&
%p_{\,1\,E\,\mu}\,=\,m (\cos(\phi),-\imath \sin(\phi) ,0_{\bot}),\,\,\,
%p_{\,2\,E\,\mu}\,=\,m\,(\cos(\phi) , \imath\sin(\phi) ,0_{\bot}),. \nonumber
%\eeqar
the transforms
\beq\label{MS151}
\D_{0}\,\rightarrow\,\imath\,\D_{4\,E}\,,\,\,\,\tv_{0}^{a}\,\rightarrow\,\imath\,\tv_{0\,E}^{a}\,
\eeq
must be performed further as well.
Therefore we obtain
\beq\label{MS16}
n_{\,1}^{\,\mu}\,\D_{\mu}\,\rightarrow\,\imath\,n_{\,+\,E}^{\mu}\,\D_{\mu\,E}\,,\,\,\,
n_{\,2}^{\,\mu}\,\D_{\mu}\,\rightarrow\,\imath\,n_{\,-\,E}^{\mu}\,\D_{\mu\,E}\,
\eeq
where
\beq \label{MS161}
%&\,&
n_{\,+\,E}^{\,\mu}\,=\,\frac{1}{\sqrt{2}}\,(1,\tan(\phi) ,0_{\bot}),\,\,\,
n_{\,-\,E}^{\,\mu}\,=\,\frac{1}{\sqrt{2}}\,(1 , -\tan(\phi) ,0_{\bot})\,
\eeq
are vectors in the Euclidean space.
Correspondingly, the Eikonal interaction term in Minkowski space can be rewritten in Euclidean space   as 
\beq\label{MS18}
-\imath\,\int\,d^{4} x \Le \Le n_{\,1}^{\,\mu}\,\D_{\mu}\,O(\tv_+)\Ra\,\D_{\bot}^{2} \Le  n_{\,2}^{\,\nu}\,\D_{\nu}\,O(\tv_-)\Ra\Ra \rightarrow
\int\,d^{4} x_{E}\Le\Le \,n_{\,+\,E}^{\mu}\,\D_{\mu\,E}\,O_{E\,+} \Ra \D_{\bot\,E}^{2} \Le n_{\,-\,E}^{\mu}\,\D_{\mu\,E}\,O_{E\,-} \Ra \Ra\,.
\eeq
with
\beq\label{MS19}
O(\tv_{\pm})\,=\,P\,e^{g\,\int_{-\infty}^{1}\, d \lambda\,(n_{\,1,2}^{\mu}\,\tv_{\mu})}\,\rightarrow\,O_{E\,\pm}\,=\,
P\,e^{\,\imath\, g\,\int_{-\infty}^{1}\,d \lambda\,( n_{\pm\,E}^{\mu}\,\tv_{\mu\,E})}\,
,\,\,\,\,\tv_{\mu\,E}\,=\,\imath\,T^{a}\,\tv_{\mu}^{a}\,.
\eeq
The remaining part of \eq{Add2} gluon's QCD Lagrangian is continued to the Euclidean space as usual.

\section{Lipatov's action in Euclidean space}

In this Section we consider the generating functional for the Lipatov's operators in the Euclidean space, here and further we omit the $E$ notation in the formulae.
We write the full Lagrangian of the approach as Lagrangian of the interacting eikonal lines  averaged over the pure YM Lagrangian obtaining the following expression:
\beqar\label{ES1}
Z[J] \,&= &\,
\frac{1}{Z^\prime} \int D\tv \, {\rm exp}\,\Big(-\, S_{YM}[\tv]\,+\,
\frac{1}{2\,g^2\,C(R)}\int\,d^{4} x\,
\Le \,n_{+}^{\mu}\,\D_{\mu}\,O_{+} \Ra\,\D_{\bot}^{2}\,\Le \,n_{-}^{\mu}\,\D_{\mu}\,O_{-}\Ra\,-\,\nonumber \\
&-&\,\frac{\imath}{2\,g\,C(R)}\int\, d^4 x \,J_{-}\,\Le \,n_{+}^{\mu}\,\D_{\mu}\,O_{+} \Ra\,-\,
\frac{\imath}{2\,g\,C(R)}\int\, d^4 x \,J_{+}\,\Le \,n_{-}^{\mu}\,\D_{\mu}\,O_{-}\Ra\,\Big)\,.
\eeqar
Introducing Lipatov's operators
\beq\label{ES2}
\T_{\pm}\,=\,\frac{1}{g}\,n_{\pm}^{\mu}\D_{\mu}\,O_{\pm}\,=\,\imath\,\Le n^{\mu}_{\,\pm}\,\tv_{\mu} \Ra\, O_{\pm}\,
\eeq
we rewrite the same generating functional as
\beqar\label{ES3}
Z[J] \,&= &\,
\frac{1}{Z^\prime} \int D\tv \, {\rm exp}\,\Big(-\, S_{YM}[\tv]\,+\,
\frac{1}{2\,C(R)}\int\,d^{4} x\,
\T_{+}\,\D_{\bot}^{2}\,\T_{-}\,-\,\nonumber \\
&-&\,\frac{\imath}{2\,C(R)}\int\, d^4 x \,J_{-}\,\T_{+}\,-\,
\frac{\imath}{2\,C(R)}\int\, d^4 x \,J_{+}\,\T_{-}\,\Big)\,.
\eeqar
Now, with the help of some auxiliary fields\footnote{In  Minkowski space these fields are Reggeon fields.} $A_{\pm}$ we obtain for the generating functional the following expression:
\beqar\label{ES4}
Z[J] \,&=&\,
\frac{1}{Z^\prime} \int D\tv \, D A\, {\rm exp}\,\Big(-\,S_{YM}[\tv]\,-\,
\frac{2}{\,C(R)}\,\int d^4 x \,A_{+}(x)\,\partial_{\bot}^{2}\,A_{-}(x)\,+\,\nonumber \\
&+&\,\frac{1}{\,C(R)}\,\int\, d^4 x \,\T_{+}\,\D^{2}_{\bot}\,A_{-}\,+\,
\frac{1}{\,C(R)}\,\int\, d^4 x \,\T_{-}\,\D^{2}_{\bot}\,A_{+}\,-
\,\frac{\imath}{\,C(R)}\,\int\, d^4 x \,\,J_{-}\,A_{+}\,-\,\nonumber \\
&-&\,
\frac{\imath}{\,C(R)}\,\int\, d^4 x \,J_{+}\,A_{-}\,+\,\frac{1}{2\,C(R)} \int d^{4}x \,J_{+}\Le\D_{\bot}^{2}\Ra^{-1} J_{-}\,\Big)\,.
\eeqar
Taking the external currents equal to zero, we write finally the generating functional for Lipatov's action in the Euclidean space:
\beqar\label{ES5}
Z[A_{+},\,A_{-}]\,&=&\,
\frac{1}{Z^\prime} \int D\tv \, {\rm exp}\,\Big(-\,S_{YM}[\tv]\,-\,
\frac{2}{\,C(R)}\,\int d^4 x \,A_{+}(x)\,\partial_{\bot}^{2}\,A_{-}(x)\,+\,\nonumber \\
&+&\,\frac{1}{\,C(R)}\,\int\, d^4 x \,\T_{+}\,\D^{2}_{\bot}\,A_{-}\,+\,
\frac{1}{\,C(R)}\,\int\, d^4 x \,\T_{-}\,\D^{2}_{\bot}\,A_{+}\,\Big)\,.
\eeqar
The classical equations of motion are usual in this case
\beq\label{ES6}
\Le\,D_{\mu}\,G^{\mu \nu}\,\Ra_{a}\,=\,\partial_{\mu}\,G_{a}^{\mu \nu}\,+\,g\,f_{abc}\tv_{\mu}^{b}\,G^{c\,\mu \nu}\,=\,j_{a}^{\nu}\,
\eeq
with only the new effective currents obtaining by the variation of the Lipatov's currents with respect to the $\tv_{4},\,\tv_{1}$ gluon fields:
\beqar\label{ES7}
j_{a}^{4}\,& = &\,-\,\frac{\imath}{N\sqrt{2}}\,
tr[\,f_{a}\,O_{+}\,f_{b}\,O^{T}_{+}\,]\,\Le  \, \D_{\bot}^{2} A_{-}^{b}\,\Ra\,-\,
\frac{\imath}{N\sqrt{2}}\,tr[\,f_{a}\,O_{-}\,f_{b}\,O^{T}_{-}\,]\,\Le \,\D_{\bot}^{2} A_{+}^{b}\Ra\, \\
j_{a}^{1}\,& = &\,-\,\frac{\imath}{N\sqrt{2}}\,
tr[\,f_{a}\,O_{+}\,f_{b}\,O^{T}_{+}\,]\,\Le \D_{\bot}^{2} A_{-}^{b} \,\Ra\,\tan (\phi)\,+\,
\frac{\imath}{N\sqrt{2}}\,tr[\,f_{a}\,O_{-}\,f_{b}\,O^{T}_{-}\,]\,\Le \D_{\bot}^{2} A_{+}^{b}\Ra\,\tan (\phi)\,.
\eeqar
Additionally, we free to redefine the auxiliary fields
\beq\label{ES8}
A_{+}\,\rightarrow\,\imath\,A_{+} / \sqrt{2}\,,\,\,\,A_{-}\,\rightarrow\,\imath\,A_{-}/ \sqrt{2}\,
\eeq
rewriting the currents as 
\beqar\label{ES9}
j_{a}^{4}\,& = &\,-\,\frac{1}{N}\,
tr[\,f_{a}\,O_{+}\,f_{b}\,O^{T}_{+}\,]\,\Le  \, \D_{\bot}^{2} A_{-}^{b}\,\Ra\,-\,
\frac{1}{N}\,tr[\,f_{a}\,O_{-}\,f_{b}\,O^{T}_{-}\,]\,\Le \,\D_{\bot}^{2} A_{+}^{b}\Ra\, \\
j_{a}^{1}\,& = &\,-\,\frac{1}{N}\,
tr[\,f_{a}\,O_{+}\,f_{b}\,O^{T}_{+}\,]\,\Le \D_{\bot}^{2} A_{-}^{b} \,\Ra\,\tan (\phi)\,+\,
\frac{1}{N}\,tr[\,f_{a}\,O_{-}\,f_{b}\,O^{T}_{-}\,]\,\Le \D_{\bot}^{2} A_{+}^{b}\Ra\,\tan (\phi)\,
\eeqar
which to LO are equal to
\beq\label{ES10}
j_{a}^{4}\,= \,\D_{\bot}^{2}\,\Le A_{-}\,+\,A_{+} \Ra\,,\,\,\,j_{a}^{1}\,= \,\D_{\bot}^{2}\,\Le A_{-}\,-\,A_{+} \Ra\,\tan(\phi)\,,
\eeq
with the corresponding change of the kinetic term of the $A_{\pm}$ fields in \eq{ES5} accounted afterwards. 

\section{The action in Euclidean "light-cone" coordinates}

 The presence of the angle in \eq{ES9}  currents determine the gluons fields as dependent on the angle through 
the equations of motion. Whereas these expressions are suitable for the analytical calculations,
the numerical implementation of any calculations can be complicated somehow because of the angle present. Therefore, we introduce the following "light-cone" coordinates in the Euclidean space.
Requiring $n_{\pm}^{\mu}\D_{\mu}\,O_{\pm}\,=\,\D_{\pm}$ we determine the "contravariant" "light-cone" coordinates as
\beq\label{LC1}
x^{+}\,=\,\frac{x^{4}\,+\,x^{1} / \tan{\phi}}{\sqrt{2}}\,,\,\,\,x^{-}\,=\,\frac{x^{4}\,-\,x^{1} / \tan{\phi}}{\sqrt{2}}\,
\eeq
and "covariant" gluon fields in the Euclidean space as
\beq\label{LC2}
\tv_{+}\,=\,\frac{\tv_{4}\,+\,\tv_{1}\,\tan{\phi}}{\sqrt{2}}\,,\,\,\,\tv_{-}\,=\,\frac{\tv^{4}\,-\,\tv^{1}\, \tan{\phi}}{\sqrt{2}}\,.
\eeq
Corresponding "covariant" and "contravariant" vectors are obtained with the help of the following metric tensor\footnote{The tensor convert form of \eq{LC1} vectors to \eq{LC2} form and vise verse,
it's action is given simply by $x^{\pm}(\phi)\,=\,x_{\pm}(\pi / 2 - \phi)$ replace.}:
\beq\label{LC3}
g_{\mu\,\nu}=\frac{1}{2\cos^{2}(\phi)}\left(
\begin{array}{c c}
1 & \cos(2\phi) \\
\cos(2\phi) & 1 \\
\end{array} \right )\,,\,
g^{\,\mu\,\nu}=\frac{1}{2\sin^{2}(\phi)}\left(
\begin{array}{c c}
1 & -\cos(2\phi) \\
-\cos(2\phi) & 1 \\
\end{array} \right )\,,\,\mu\,\nu = +\,-\,.
\eeq
In this case we have for the Lagrangian:
\beq\label{LC4}
L_{QCD}\,=\,\frac{1}{2}\, G_{+ -}^{a}\,G^{+ -}_{a}\,+\,\frac{1}{2}\, G_{+ i}^{a}\,G^{+ i}_{a}\,+
\frac{1}{2}\, G_{- i}^{a}\,G^{- i}_{a}\,+\,\frac{1}{4} \, G_{i j}^{a}\,G^{i j}_{a}\,
\eeq
and for the effective currents \eq{MS19}:
\beq\label{LC5}
O_{+}\,=\,e^{\,\imath\, g\,\int_{-\infty}^{x^{+}}\,d x^{'\,+}\,\tv_{+}(x^{'\,+},\,x^{-},\,x_{\bot})}\,,\,\,\,
O_{-}\,=\,e^{\,\imath\, g\,\int_{-\infty}^{x^{-}}\,d x^{'\,-}\,\tv_{-}(x^{+},\,x^{'\,-},\,x_{\bot})}\,.
\eeq
We see, therefore, that in these "light-cone" coordinates the \eq{ES5} effective action does not contain the angle, it must be accounted only once in the \eq{LC2} definition of "covariant"
and corresponding "contravariant" gluon fields. The price for that is the doubled number of the longitudinal gluon fields which are depend each on other trough the 
transformations with the use of \eq{LC3} metric tensor. Additional advantage of the introduced coordinates is that we can use here one from the $\tv_{\pm}\,=\,0$ gauges that  simplifies
the structure of the \eq{LC5} currents terms and corresponding numerical calculations. 

\section{The action in axial gauge}

 The "light-cone" coordinates introduced above are not so suitable for the analytical calculation. Eliminating the angle's dependence of the effective currents, the new gluons fields "move"
the angle in the 
l.h.s. of \eq{ES6} written in the terms of only "covariant" coordinates. The Lagrangian \eq{LC4} does not help in this case, there is no simple rule which allows to raise and lower the corresponding indexes with the help of \eq{LC3} tensor. Therefore, we consider the analytical solution of \eq{ES6} equations of motion taking $A_{1}\,=\,0$ axial gauge, where in this case the condition
$O_{+}\,=\,O_{-}$  also is satisfied.
We have correspondingly to LO precision\footnote{We omit here color indexes of the fields for the shortness of the notations.}:
\beqar\label{AG1}
&\,&\,\D_{4}^{2}\,\tv_{i}\,+\,\D_{1}^{2}\,\tv_{i}\,-\,\D_{4}\,\D_{i}\,\tv_{4}\,=\,0\,\\
&\,&\,\D_{1}^{2}\,\tv_{4}\,+\,\D_{i}^{2}\,\tv_{4}\,-\,\D_{4}\,\D_{i}\,\tv_{i}\,=\,j_{4}\,\\
&\,&\,-\,\D_{4}\,\D_{1}\tv_{4}\,-\,\D_{i}\,\D_{1}\,\tv_{i}\,=\,j_{1}\,.
\eeqar
The solution of the equations are the following functions:
\beq\label{AG2}
\tv_{4}^{cl}(A_{+},\,A_{-})\,=\,\Box^{-1}\,\Le j_{4}\,-\,\D_{4}\,\D_{1}^{-1}\,j_{1} \Ra\,,\,\,\,
\tv_{i}^{cl}(A_{+},\,A_{-})\,=\,-\D_{i}\,\Box^{-1}\,\Le \D_{1}^{-1}\,j_{1} \Ra\,.
\eeq
The third equation from the system for the two unknown functions is the condition of transversality of the currents:
\beq\label{AG3}
\D_{\mu}\,j_{\mu}\,=\,0
\eeq
that to LO can be written with the help of \eq{ES10} as:
\beq\label{AG4}
\D_{+}\,A_{-}\,=\,\D_{-}\,A_{+}\,=\,0\,\rightarrow\,A_{-}\,=\,A_{-}(x^{-})\,,\,\,A_{+}\,=\,A_{+}(x^{+})\,,
\eeq
see definitions in the above section. Now we ready to incorporate into the effective action framework the classical instanton solution for the gluon fields. 
Writing  the gluon fields in Euclidean space as  classical solution plus fluctuations around it
\beq\label{AG5}
\tv_{4}\,=\,v_{4}^{cl}\,+\,\varepsilon_{4}\,=\,\tv_{4}^{inst}\,+\,\tv_{4}^{cl}(A_{+},\,A_{-})\,+\,\varepsilon_{4}\,,\,\,\,
\tv_{i}\,=\,v_{i}^{cl}\,+\,\varepsilon_{i}\,=\,\tv_{i}^{inst}\,+\,\tv_{i}^{cl}(A_{+},\,A_{-})\,+\,\varepsilon_{i}\,,
\eeq
we will obtain for \eq{ES5} generating functional
\beqar\label{AG6}
Z[\tv^{inst},\,A_{+},\,A_{-}] \,&=&\,
\frac{1}{Z^\prime} \int D\varepsilon \,  {\rm exp}\,\Big(-\,S_{YM}[\tv]\,-\,
\frac{2}{\,C(R)}\,\int d^4 x \,A_{+}(x)\,\partial_{\bot}^{2}\,A_{-}(x)\,+\,\nonumber \\
&+&\,\frac{1}{\,C(R)}\,\int\, d^4 x \,\T_{+}\,\D^{2}_{\bot}\,A_{-}\,+\,
\frac{1}{\,C(R)}\,\int\, d^4 x \,\T_{-}\,\D^{2}_{\bot}\,A_{+}\,\Big)\,.
\eeqar
The obtained functional determines the vertices of interactions of $A_{\pm}$ fields with the instanton fields in the framework of high energy Euclidean QCD RFT, that after the inverse continuation to Minkowski space
will determine the vertices of interactions of Reggeon  with instanton fields as well.

 We also note, that using the diagrammatic approach of \cite{LipatovEff1,EffAct}, the effective currents determine the Feynman rules for the construction of the
vertices of interaction of gluons with instanton and Reggeon fields. Namely, instead the \eq{AG5} representation of gluon fields, the any interaction vertex of interest can be constructed by the 
$\tv\,\rightarrow\,\tv\,+\,\tv^{inst}$ substitution performed directly in the 
effective currents and their consequent expansion into the perturbative series similarly to done in \cite{LipatovEff1}.

\section{Conclusion}

 In this note we clarified two issues concerning the formalism of  high energy QCD Lipatov's effective action.
Namely, we expanded the formalism to Euclidean space having in mind the following possible applications of the Euclidean version of the action. 
 
 The first one is the numerical (lattice) calculation in the framework with the Euclidean action. With the help of \eq{ES5} generating functional any correlator of $A_{\pm}$
fields can be calculated\footnote{For the color correlators an additional regularization of the effective currents must be introduced.}. In Minkowski space it will allow to trace the high energy behavior 
of the arbitrary Reggeon's correlators. For example, taking BFKL colorless state we can calculate the correlator in Euclidean state for the different values of $\phi$ angle. The reverse continuation 
to Minkowski space, therefore, will allow to interpolate the behavior of the correlator as function of energy on the base of the points obtained in the Euclidean space. These non-perturbative calculations of the high energy asymptotic behavior of the  Pomeron
(and other correlators)
with the unitarity corrections included is an interesting task due the importance of the BFKL calculus in the high-energy QCD. Another interesting possibility of the application of the formalism is the 
connection of the Wilson lines correlators and correlators of $A_{\pm}$ fields, see \eq{ES3}-\eq{ES4}, the knowledge of  $A_{\pm}$ correlators will determine the correlators of Wilson lines as well.

 Another application of the Euclidean version of the action, is that it determines the correct interaction vertices of the correlators of $A_{\pm}$ fields with the instanton fields.
Namely, \eq{AG6}, after the integration with respect to the classical instanton fields, will provide instanton induced corrections to the Reggeon fields correlators, i.e. to the propagator of reggeized gluons, BFKL Pomeron, et cetera. These corrections are interesting to account, see \cite{Inst} for the different applications of the instanton contributions in the high energy scattering processes.
The calculations related to this especially interesting task we plan to begin as the next step in the development of the framework.

\end{document}